# Quasi-one-dimensional ballistic ring in crossed high-frequency electric fields

E.M. Epshtein[1], E.G. Fedorov[2], G.M. Shmelev[3]

[1] Institute of Radio Engineering and Electronics of the Russian Academy of Sciences, Fryazino, Russia; e-mail: eme253@ms.ire.rssi.ru

[2] Volgograd State Architectural & Engineering University, Volgograd, Russia; e-mail: eduard-f@mail.ru

[3] Volgograd State Pedagogical University, Volgograd, Russia; e-mail: shmelev@fizmat.vspu.ru

We study electron dynamics in a quasi-one-dimensional ballistic ring driven by two crossed high-frequency electric fields parallel to the ring plane. The averaged dipole moment and emission intensity are calculated. The emission polarization coincides with the direction of one of the fields. A possibility is shown of the polarization switching to perpendicular direction under changes in the field amplitudes and frequencies.

The mesoscopic structure technology stimulates both experimental and theoretical investigations of low-dimensional systems, specifically, quasi-one-dimensional rings [1 – 4]. The most of the works on this problem are dedicated to the quantum phenomena. However, the quasi-one-dimensional rings have interesting classical electrodynamic properties [5 – 11]. In present paper, we investigate the quasi-one-dimensional ring response to two plane electromagnetic waves polarized perpendicular to each other which propagate along the normal to the ring plane.

Consider a planar ring that is a quantum well between two concentric potential barriers. The ring width is small compared to its radius $R$. The electron motion is quantized along the ring radius, while it is classical along the circumference. Assume that the electron mean free path is large compared to $2\pi R$, so that electrons in the ring move ballistically.

Let two waves mentioned are incident on the ring with the wavelengths greater then the ring diameter. In the dipole approximation, only electric field of the waves $\boldsymbol{E} = \{E_1 \sin(\omega_1 t + \beta_1), E_2 \sin(\omega_2 t + \beta_2)\}$ acts on the electrons. Assume that before the fields turn on (at $t = 0$) the electron has energy $W$ that is energy of rotational motion along the ring circumference. We suppose that the electrons interact only with the external fields and consider single-electron dynamics [5 – 11]. The electron motion in the ring obeys the following equation:

$$\ddot{\varphi}+\Omega_1^2 \sin\varphi\sin(\omega_1 t+\alpha_1)-\Omega_2^2 \cos\varphi\sin(\omega_2 t+\alpha_2)=0 \tag{1}$$

where $\Omega_1^2 = |e|E_1/(mR)$, $\Omega_2^2 = |e|E_2/(mR)$, $\varphi$ is the angular coordinate, $\alpha_{1,2} = \beta_{1,2} \pm \pi$.

To determine slow component of the electron motion, we use the Kapitza averaging method [12], in that the angular coordinate is presented as $\varphi = \Phi + \xi$, where $\Phi$ and $\xi$ describe slow motion and small high-frequency oscillations, respectively ($|\xi| \ll |\Phi|$). The slow motion is described by equation $mR^2\ddot{\Phi} = -\partial U_{eff}/\partial \Phi$ where

$$U_{eff} = mR^2 \left\{ \frac{1}{4}\left[\left(\frac{\Omega_1^2}{\omega_1}\right)^2 - \left(\frac{\Omega_2^2}{\omega_2}\right)^2\right]\sin^2\Phi + \frac{(\Omega_1\Omega_2)^2}{2}\langle\sin(\omega_1 t+\alpha_1)\sin(\omega_2 t+\alpha_2)\rangle \times \right.$$

$$\left. \times \left[\Phi\cdot\left(\frac{1}{\omega_1^2}-\frac{1}{\omega_2^2}\right)-\frac{1}{2}\left(\frac{1}{\omega_1^2}+\frac{1}{\omega_2^2}\right)\sin 2\Phi\right]\right\}, \tag{2}$$

the angular brackets $\langle\ldots\rangle$ means averaging over fast periods.

Let us consider some specific cases.

1. At $\omega_1 \neq \omega_2$ and/or $\alpha_1 - \alpha_2 = \pm\pi/2$, we obtain from Eq. (2)

$$U_{eff} = \frac{mR^2}{4}\left[\left(\frac{\Omega_1^2}{\omega_1}\right)^2 - \left(\frac{\Omega_2^2}{\omega_2}\right)^2\right]\sin^2\Phi. \tag{3}$$

At $\Omega_1^2/\omega_1 > \Omega_2^2/\omega_2$, effective potential energy (3) has minima at $\Phi = 0$ and $\Phi = \pi$, while at $\Omega_1^2/\omega_1 < \Omega_2^2/\omega_2$, the minima are at $\Phi = \pm\pi/2$. In the case of elliptic polarization ($\omega_1 = \omega_2 = \omega, \alpha_1 - \alpha_2 = \pm\pi/2$), Eq. (3) takes form of

$$U_{eff} = \frac{mR^2}{4\omega^2}\left[(\Omega_1^2)^2 - (\Omega_2^2)^2\right]\sin^2\Phi. \tag{4}$$

2. If $\omega_1 = \omega_2 = \omega$, then

$$U_{eff} = \frac{mR^2}{4\omega^2}\left[(\Omega_1^4 - \Omega_2^4)\sin^2\Phi - (\Omega_1\Omega_2)^2 \cos(\alpha_1-\alpha_2)\sin 2\Phi\right]. \tag{5}$$

At $\Omega_1 = \Omega_2 = \Omega$,

$$U_{eff} = -\frac{mR^2}{4\omega^2}\Omega^4 \cos(\alpha_1-\alpha_2)\sin 2\Phi. \tag{6}$$

The potential energy (6) has minima $\Phi = \pi/4$ and $\Phi = 5\pi/4$.

3. If $\Omega_1^2/\omega_1 = \Omega_2^2/\omega_2$, then it follows from Eq. (3) that $U_{eff} = 0$ under conditions $\omega_1 \neq \omega_2$ and/or $\alpha_1 - \alpha_2 = \pm \pi/2$.

From the potential energy (3) the following equation is obtained for the averaged electron motion:

$$\ddot{\Phi} + \frac{1}{4}\left[\left(\frac{\Omega_1^2}{\omega_1}\right)^2 - \left(\frac{\Omega_2^2}{\omega_2}\right)^2\right]\sin 2\Phi = 0. \tag{7}$$

The dipole moment (relative to the ring center) and the emission intensity per electron are defined as [13]

$$\boldsymbol{P} = eR\{\cos\Phi, \sin\Phi\}, \tag{8}$$

$$I = \frac{2}{3c^3}\ddot{\boldsymbol{P}}^2 = \frac{2}{3c^3}(eR)^2(\dot{\Phi}^4 + \ddot{\Phi}^2). \tag{9}$$

Solution of Eq. (7) with initial conditions $\Phi_0 = \Phi(0)$, $\dot{\Phi}_0 = \dot{\Phi}(0) = R^{-1}(2W/m)^{1/2}$ is

$$\Phi = \Psi + \lambda, \tag{10}$$

$$\Psi = \begin{cases} \arcsin\{q\,\text{sn}[\omega_0 t + F(\arcsin(q^{-1}\sin(\Phi_0 - \lambda)), q), q]\}, & q < 1, \\ \arcsin\{\tanh[\omega_0 t + F(\Phi_0 - \lambda, 1)]\}, & q = 1, \\ \arcsin\{\text{sn}[q\omega_0 t + F(\Phi_0 - \lambda, q^{-1}), q^{-1}]\}, & q > 1, \end{cases} \tag{11}$$

where $\lambda = 0$ at $\Omega_1^2/\omega_1 > \Omega_2^2/\omega_2$ and $\lambda = \pi/2$ at $\Omega_1^2/\omega_1 < \Omega_2^2/\omega_2$,

$$\omega_0 = \sqrt{\frac{1}{2}\left|\left(\frac{\Omega_1^2}{\omega_1}\right)^2 - \left(\frac{\Omega_2^2}{\omega_2}\right)^2\right|}, \tag{12}$$

$$q^2 = \frac{\dot{\Phi}_0^2}{\omega_0^2} + \frac{1}{2}(1 - \cos 2(\Phi_0 - \lambda)), \tag{13}$$

sn($x$, $k$) is the elliptic sine, $F(x, k)$ is the incomplete elliptic integral of the first kind.

The fundamental frequency of the electron slow motion is

$$\omega_{slow} = \omega_0 \frac{\pi}{2}\begin{cases} \dfrac{1}{K(q)}, & q < 1, \\ 0, & q = 1, \\ \dfrac{q}{K(q^{-1})}, & q > 1, \end{cases} \tag{14}$$

where $K(k)$ is the complete elliptic integral of the first kind [14].

From Eqs. (9) and (10) we find the emission intensity from the

$$\text{ring: } I = \frac{2}{3c^3}(eR)^2 \omega_0^4 \begin{cases} q^4 + q^2(1-2q^2)\text{sn}^2[\omega_0 t + F(\arcsin(q^{-1}\sin(\Phi_0 - \lambda)), q), q], & q < 1, \\ \text{sech}^2[\omega_0 t + F(\Phi_0 - \lambda, 1)], & q = 1, \\ q^4 + (1-2q^2)\text{sn}^2[q\omega_0 t + F(\Phi_0 - \lambda, q^{-1}), q^{-1}], & q > 1. \end{cases} \quad (15)$$

The mean emission intensity is

$$\langle I \rangle = \frac{2}{3c^3}(eR)^2 \omega_0^4 G(q), \quad (16)$$

where $G(q)$ takes form of

$$G(q) = \begin{cases} q^4 + (1-2q^2)\left[1 - \frac{E(q)}{K(q)}\right], & q < 1, \\ 0, & q = 1, \\ q^4 + q^2(1-2q^2)\left[1 - \frac{E(q^{-1})}{K(q^{-1})}\right], & q > 1, \end{cases} \quad (17)$$

where $E(k)$ is the complete elliptic integral of the second kind [14].

At $q \ll 1$ the electron executes field-driven small oscillations. In that case, $G(q) \cong 4/3$, so that

$$\langle I \rangle = \frac{8}{9c^3}(eR)^2 \omega_0^4. \quad (18)$$

At $\omega_0 \ll \dot{\Phi}_0$ (e.g., at $E_1/\omega_1 \approx E_2/\omega_2$) we have $q^2 \cong \dot{\Phi}_0^2/\omega_0^2 \gg 1$, so that $G(q) \cong q^4$, $(\omega_0 q)^4 \cong \dot{\Phi}_0^4$, and

$$\langle I \rangle = I_0 = \frac{2}{3c^3}(eR)^2 \dot{\Phi}_0^4, \quad (19)$$

that coincides with the expression for an electron moving along the ring circumference with constant angular velocity [13] corresponding to energy $W = R^2 \dot{\Phi}_0^2 m/2$. That case corresponds, in particular, to a circularly polarized electromagnetic wave ($\alpha_1 - \alpha_2 = \pm \pi/2$, $E_1 = E_2 = E_0$, $\omega_1 = \omega_2 = \omega$), when $U_{\text{eff}} = 0$. We intend to publish elsewhere detailed calculations of the emission from a ballistic ring in a circularly polarized electromagnetic wave.

In analyzing the dependence $\langle I \rangle$ on $q$, we consider $q$ as a function of $\dot{\Phi}_0^2/\omega_0^2$ under fixed $\Phi_0$ or as a function of $\Phi_0$ under fixed $\dot{\Phi}_0^2/\omega_0^2$.

Let $q = q(\dot{\Phi}_0^2/\omega_0^2)$ under fixed $\Phi_0$. Then it follows from Eq. (16)

$$\langle I \rangle = I_0 \frac{G(q)}{q^4}\left[1 + M(q) + \frac{1}{4}M^2(q)\right], \quad (20)$$

where $M(q) = \dfrac{(1 - \cos 2(\Phi_0 - \lambda))}{q^2 - (1/2)(1 - \cos 2(\Phi_0 - \lambda))}$.

If $q = q(\Phi_0)$ under fixed $\dot{\Phi}_0^2/\omega_0^2$, then

$$\langle I \rangle = I_0 \frac{G(q)}{q^4}\left[1 + B(q) + \frac{1}{4}B^2(q)\right], \quad (21)$$

where $B(q) = 2(q^2 \omega_0^2/\dot{\Phi}_0^2 - 1)$.

The dependencies of $\langle I \rangle/I_0$ on $q$ under $\Omega_1^2/\omega_1 > \Omega_2^2/\omega_2$ condition described by Eqs. (20) and (21) are shown in Figs. 1 and 2, respectively.

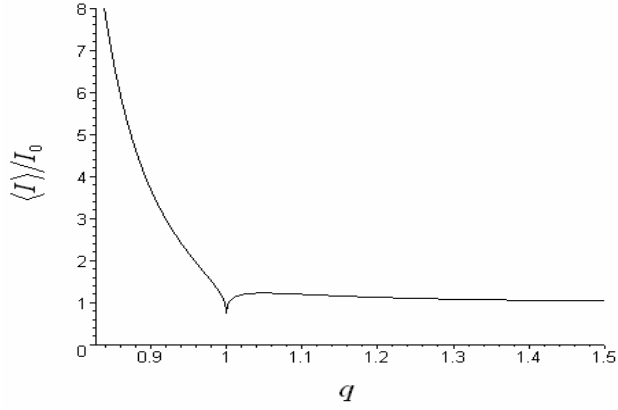

Fig. 1. The dependence of $\langle I \rangle/I_0$ on $q = q(\dot{\Phi}_0^2/\omega_0^2)$ at $\Phi_0 = \pi/4$.

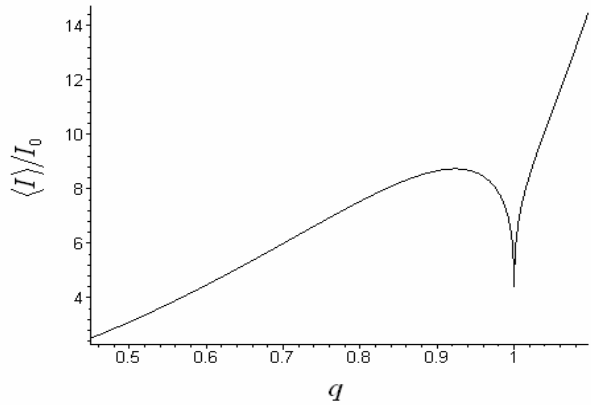

Fig. 2. The dependence of $\langle I \rangle/I_0$ on $q = q(\Phi_0)$ at $\dot{\Phi}_0/\omega_0 = 0.45$.

Note that a situation is possible when $\langle I \rangle / I_0 \gg 1$, that is the emission intensity, corresponding to the electron free rotation in the ring, is small compared to the wave-driven one.

To find the ring mean dipole moment $\langle \mathbf{P} \rangle$, we substitute (10) in (8) and average over the period $T_{slow} = 2\pi/\omega_{slow}$. At $q > 1$, always $\langle \mathbf{P} \rangle = \{0, 0\}$. At $q < 1$, two case realize, $\langle \mathbf{P} \rangle = eR\left\{\dfrac{\pi}{2\mathbf{K}(q)}, 0\right\}$ under $\Omega_1^2/\omega_1 > \Omega_2^2/\omega_2$ condition and $\langle \mathbf{P} \rangle = eR\left\{0, \dfrac{\pi}{2\mathbf{K}(q)}\right\}$ under $\Omega_1^2/\omega_1 < \Omega_2^2/\omega_2$ condition. Therefore, the emission polarization direction coincides with the polarization of one of two incident waves. Changing the relation between $\Omega_1^2/\omega_1$ and $\Omega_2^2/\omega_2$, it can be possible to switch the ring emission polarization.

If the ring contains $N$ electrons, then the expressions for $\mathbf{P}$ and $\langle \mathbf{P} \rangle$ should be multiplied by $N$, while the expressions for $I$ and $\langle I \rangle$ should be multiplied by $N^2$.

Let us make some estimates. At $R = 5 \times 10^{-5}$ cm, $m = 0{,}1 m_e$, $W = 2 \times 10^{-3}$ eV, $\omega_1 = 1 \times 10^{13}$ s$^{-1}$, $\omega_2 = 1 \times 10^{14}$ s$^{-1}$, $E_1 = 1.5 \times 10^4$ V/cm, $E_2 = 3 \times 10^3$ V/cm, we have $\Omega_1 \approx 2.3 \times 10^{12}$ s$^{-1}$, $\Omega_2 \approx 1.0 \times 10^{12}$ s$^{-1}$, $\omega_0 \approx 3.7 \times 10^{11}$ s$^{-1}$, $\dot{\Phi}_0/\omega_0 \approx 0.45$, $\Omega_1^2/\omega_1 > \Omega_2^2/\omega_2$ ($\lambda = 0$, see Eq. (10)). At $\Phi_0 = \pi/4$ and indicated values of the other parameters, we obtain $q \approx 0.8$ (see Eq. (13)), $\omega_{slow} \approx 2.8 \times 10^{11}$ s$^{-1}$ (see (14)). Therefore, it follows from (20) or (21) that $\langle I \rangle / I_0 \approx 8.1$. The $\omega_{1,2} \gg \omega_{slow}$ condition necessary to apply the Kapitza method, fulfils at chosen values of the wave frequencies. At the electron collision frequency $\nu = 5 \times 10^9$ s$^{-1}$, we have $2\pi\nu \ll \omega_{slow}$ that justifies using the collisionless approximation for the electron dynamics in the ring.

The work was supported partly by the Russian Foundation on Basic Research.